\documentstyle[12pt,aaspp4,psfig,natbib]{article}

\newcommand{\km}{km s$^{-1}$}
\newcommand{\Ly}{Ly$\alpha$}

\def\gtorder{\mathrel{\raise.3ex\hbox{$>$}\mkern-14mu
             \lower0.6ex\hbox{$\sim$}}}
\def\ltorder{\mathrel{\raise.3ex\hbox{$<$}\mkern-14mu
             \lower0.6ex\hbox{$\sim$}}}
\def\proptwid{\mathrel{\raise.3ex\hbox{$\propto$}\mkern-14mu
             \lower0.6ex\hbox{$\sim$}}}
\def\arcsec{\ifmmode '' \else $''$\fi}

\def\arcsecpoint{\ifmmode ''\!. \else $''\!.$\fi}

\def\kms{\ifmmode {\rm km\ s}^{-1} \else km s$^{-1}$\fi}
\def\Msun{\ifmmode {\rm M}_{\odot} \else M$_{\odot}$\fi}
\def\Lsun{\ifmmode {\rm L}_{\odot} \else L$_{\odot}$\fi}
\def\Zsun{\ifmmode {\rm Z}_{\odot} \else Z$_{\odot}$\fi}

\def\ergscm2{ergs\,s$^{-1}$\,cm$^{-2}$}
\def\qo{\ifmmode q_{\rm o} \else $q_{\rm o}$\fi}
\def\Ho{\ifmmode H_{\rm o} \else $H_{\rm o}$\fi}
\def\ho{\ifmmode h_{\rm o} \else $h_{\rm o}$\fi}

\def\vFWHM{\ifmmode v_{\mbox{\tiny FWHM}} \else
            $v_{\mbox{\tiny FWHM}}$\fi}
\def\CCF{\ifmmode F_{\it CCF} \else $F_{\it CCF}$\fi}
\def\ACF{\ifmmode F_{\it ACF} \else $F_{\it ACF}$\fi}
\def\Halpha{\ifmmode {\rm H}\alpha \else H$\alpha$\fi}
\def\Hbeta{\ifmmode {\rm H}\beta \else H$\beta$\fi}
\def\Hgamma{\ifmmode {\rm H}\gamma \else H$\gamma$\fi}
\def\Hdelta{\ifmmode {\rm H}\delta \else H$\delta$\fi}
\def\Lya{\ifmmode {\rm Ly}\alpha \else Ly$\alpha$\fi}
\def\Lyb{\ifmmode {\rm Ly}\beta \else Ly$\beta$\fi}
\def\Lyg{\ifmmode {\rm Ly}\beta \else Ly$\gamma$\fi}

\def\cii{C\,{\sc ii}}
\def\ciii{\ifmmode {\rm C}\,{\sc iii} \else C\,{\sc iii}\fi}
\def\civ{\ifmmode {\rm C}\,{\sc iv} \else C\,{\sc iv}\fi}

\def\niii{N\,{\sc iii}}

\def\nv{N\,{\sc v}}

\def\oiii{O\,{\sc iii}}
\def\o5007{[O\,{\sc iii}]\,$\lambda5007$}
\def\oiv{O\,{\sc iv}}

\def\siiv{Si\,{\sc iv}}

\def\o{\o}
\begin{document}
%\vspace{-1in}
\title{WHAT DETERMINES THE DEPTH OF BALS? \\
       KECK HIRES OBSERVATIONS OF BALQSO 1603+3002}

\author{Nahum Arav\footnote{IGPP
                LLNL, L-413, P.O. Box 808, Livermore, CA 94550;  I:
                narav@igpp.llnl.gov}, 
Robert H. Becker$^{1,}$\footnote{Physics Department,
 University of California, Davis, CA 95616},
Sally A. Laurent-Muehleisen$^{1,2}$, Michael D. Gregg$^{1,2}$, \\
Richard L. 
White\footnote{Space Telescope Science Institute, Baltimore, MD 21218} 
and Martijn de Kool\footnote{Astrophysical Theory Centre, 
John Dedman Building, ANU ACT
0200, Australia}}

%\today  

In press with the Astrophysical Journal. 

\begin{abstract}          

We find that the depth and shape of the broad absorption lines (BALs)
in BALQSO 1603+3002 are determined largely by the fraction of the
emitting source which is covered by the BAL flow.  In addition, the
observed depth of the BALs is poorly correlated with their real
optical depth.  The implication of this result is that abundance
studies based on direct extraction of column densities from the depth
of the absorption troughs are unreliable.  Our conclusion is based on
analysis of unblended absorption features of two lines from the same
ion (in this case the \siiv\ doublet), which allows unambiguous
separation of covering factor and optical depth effects.  The complex
morphology of the covering factor as a function of velocity suggests
that the BALs are produced by several physically separated outflows.
The covering factor is ion dependent in both depth and velocity width.
We also find evidence that in BALQSO 1603+3002 the flow does not cover
the broad emission line region.

{\it Subject headings:} quasars: absorption lines

\end{abstract}          
\setcounter{footnote}{0}
\section{INTRODUCTION}

Broad Absorption Line (BAL) QSOs are a manifestation of AGN
outflows.  BALs are associated with prominent resonance lines such as
\civ~$\lambda$1549, \siiv~$\lambda$1397, \nv~$\lambda$1240, and \Ly\
$\lambda$1215.  They appear in about 10\% of all quasars \cite[]{foltz90}
 and have  typical velocity widths of $\sim10,000$ \km\ 
 \cite[]{wtc,turnshek88a} and terminal velocities
of up to 50,000 \km.  The small percentage of BALQSOs among quasars is
generally interpreted as an orientation effect \cite[]{Weymann91}
and it is probable that the majority of quasars and other types of AGN
harbor intrinsic outflows.

A crucial issue in the study of the outflows is whether the observed
depth of the BALs is determined by the column density along the line
of sight, or is due to `non-black saturation' --- 
the partial covering of the emission source by an
optically thick flow.  Non-black saturation can also be
caused by filling in the bottom of the troughs by scattered photons.
The question of column density vs. geometry (i.e., covering factor) is
especially important for determining the ionization equilibrium and
abundances (IEA) of the BAL material.  Inferences about the IEA in the
BAL region are made by simulating BAL ionic-column-densities
($N_{ion}$) using photoionization codes.  Several groups
\cite[]{korista96,turnshek96,hamann96s} have used extracted $N_{ion}$
from HST observations of BALQSO 0226--1024 \cite[]{k92} in their IEA
studies while introducing innovative theoretical approaches to the
problem. These studies, however, used the BAL {\it apparent} optical
depths (defined as $\tau=-ln(I)$, where $I$ is the residual intensity
seen in the trough) to determine $N_{ion}$.  The hazard of this
approach is that the {\it apparent} optical depths in the BALs cannot
be directly translated to realistic $N_{ion}$ unless the covering
factor and level of saturation are known.  In saturated BALs the
inferred {\it apparent} $N_{ion}$ are only lower limits to the real
$N_{ion}$, making conclusions regarding IEA in BALQSOs, such as very
high BAL metallicity \cite[]{turnshek96}, highly uncertain.

Recently, several groups presented evidence for non-black saturation
in BALs \cite[]{arav97me1,arav99a,barlow97,telfer98}; however, the
importance of the phenomenon and its detailed study as a function of
velocity across the absorption troughs are still in the preliminary
stages.  Here we present such a study of Keck HIRES observations of
BALQSO 1603+3002 ($z=2.03$).  The source was discovered during the
FIRST (Faint Images of the Radio Sky at Twenty centimeters) Bright
Quasar Survey (FBQS, Gregg et al.\ 1996; White et al.\ 1999),
\nocite{gregg96,white99} which selects quasar candidates by comparing
the catalog of radio sources found by the VLA FIRST survey
\cite[]{becker95,white97} with the APM catalog of the POSS-I plates
\cite[]{mcmahon92}.  One of the biggest surprises from the FBQS is the
prevalence of BAL quasars in this radio-selected sample. Although
previous studies indicate that none of the known BAL quasars are
radio-loud, BAL quasars have been found in the FBQS at a rate equal to
or greater than that for optically-selected quasar samples. This
result motivated us to begin an in depth study of the BAL quasars in
the FBQS. In this paper we present a high-resolution spectrum of FIRST
J160354.2+300209 (hereafter BALQSO 1603+3002).  The low-resolution
discovery spectrum is presented in \nocite{white99} (White et al.\
1999). BALQSO 1603+3002 is a radio-loud quasar with a flux density of
54 mJy at 1400 MHz and an optical magnitude of B=18.0
 [$\log(R^*)=2.01$].  In \S~4 we establish
that the absorption in this object is BAL in nature.  BALQSO 1603+3002
is a high ionization BALQSO, in contrast to the two previously
published BALs from the FIRST survey \cite[]{becker97}. In this paper
we will focus on the optical properties of this object.

\section{ANALYSIS}

\subsection{Data Acquisition and Reduction}

On May 18, 1998 we used the High Resolution Echelle Spectrometer
\nocite{vogt94} (HIRES, Vogt et al.\ 1994) on the Keck-1 10-m
telescope to obtain three 40 minute exposures of BALQSO 1603+3002
covering 3900 -- 6000 \AA\ using a 1\farcs1 wide slit.  The orders
overlapped up to 5128~\AA, beyond which small gaps occur between
orders.  The slit was rotated to the parallactic angle to minimize
losses due to differential atmospheric refraction.  The observing
conditions were excellent with subarcsecomd seeing and near
photometric transparency. The spectra were extracted using routines
tailored for HIRES reductions \cite[]{barlow99}, normalizing the
continuum to unity.  The resolution of the extracted spectrum varies
from 3.4 to 3.6 pixels FWHM, being 0.119\AA\ at 5000\AA\ or 6.5 \km\
in velocity space.

The continuum signal-to-noise of the extracted data is roughly 10-15
per pixel.  In the analysis which follows, we boxcar smoothed the
spectrum by 10 pixels, increasing the continuum signal-to-noise to
30-50 throughout the wavelength regions of interest while retaining
sufficient velocity resolution for our analysis ($\sim18$~\km).

\clearpage

\begin{figure}[t]
  \vspace*{-0.6cm}
  \begin{minipage}[t]{9.cm}
\hspace*{9.cm}
    \psfig{file=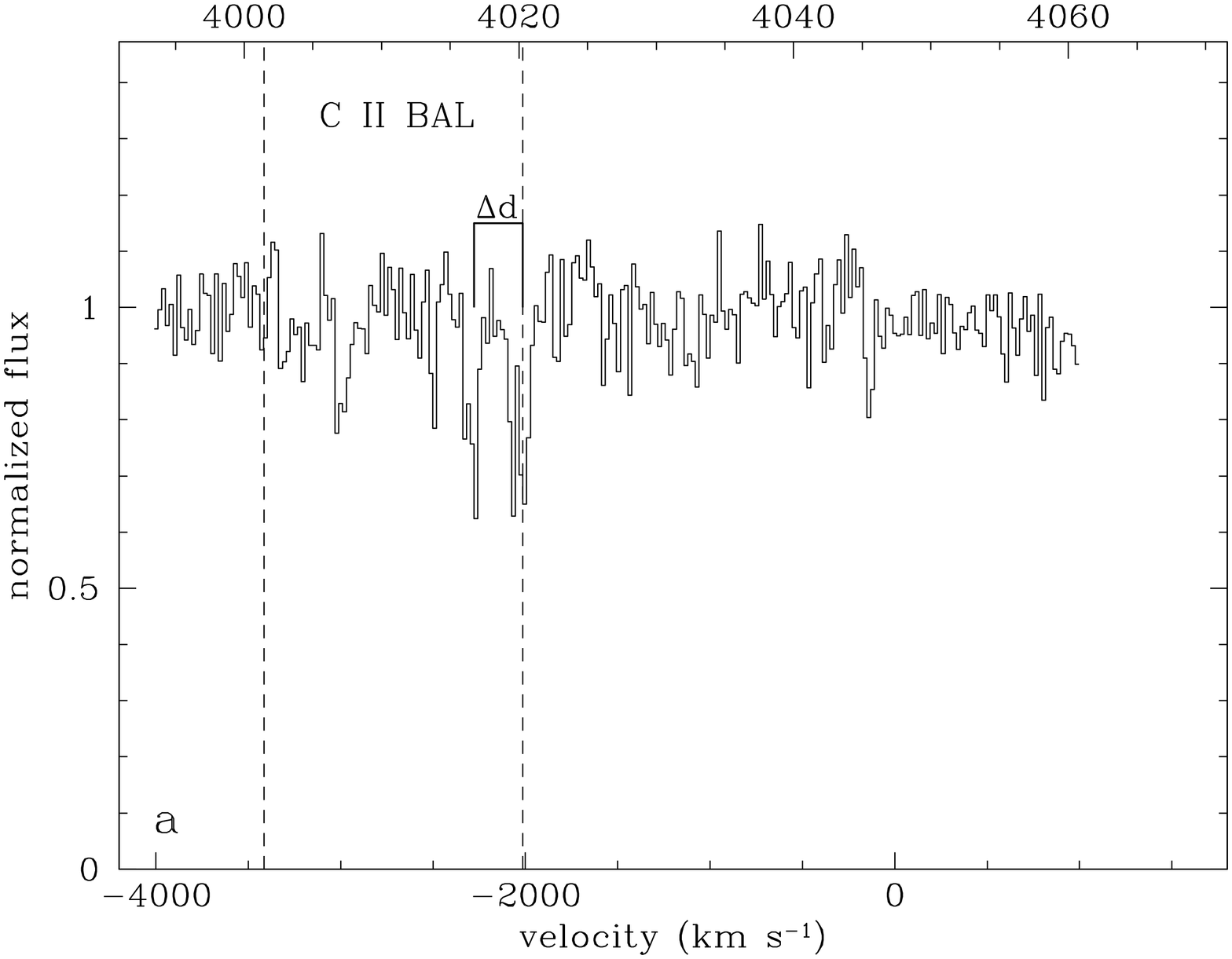,angle=-90,height=6.0cm,width=9.cm}
  \end{minipage}\hfill
  \begin{minipage}[t]{9.cm}
\hspace*{9.cm}
    \psfig{file=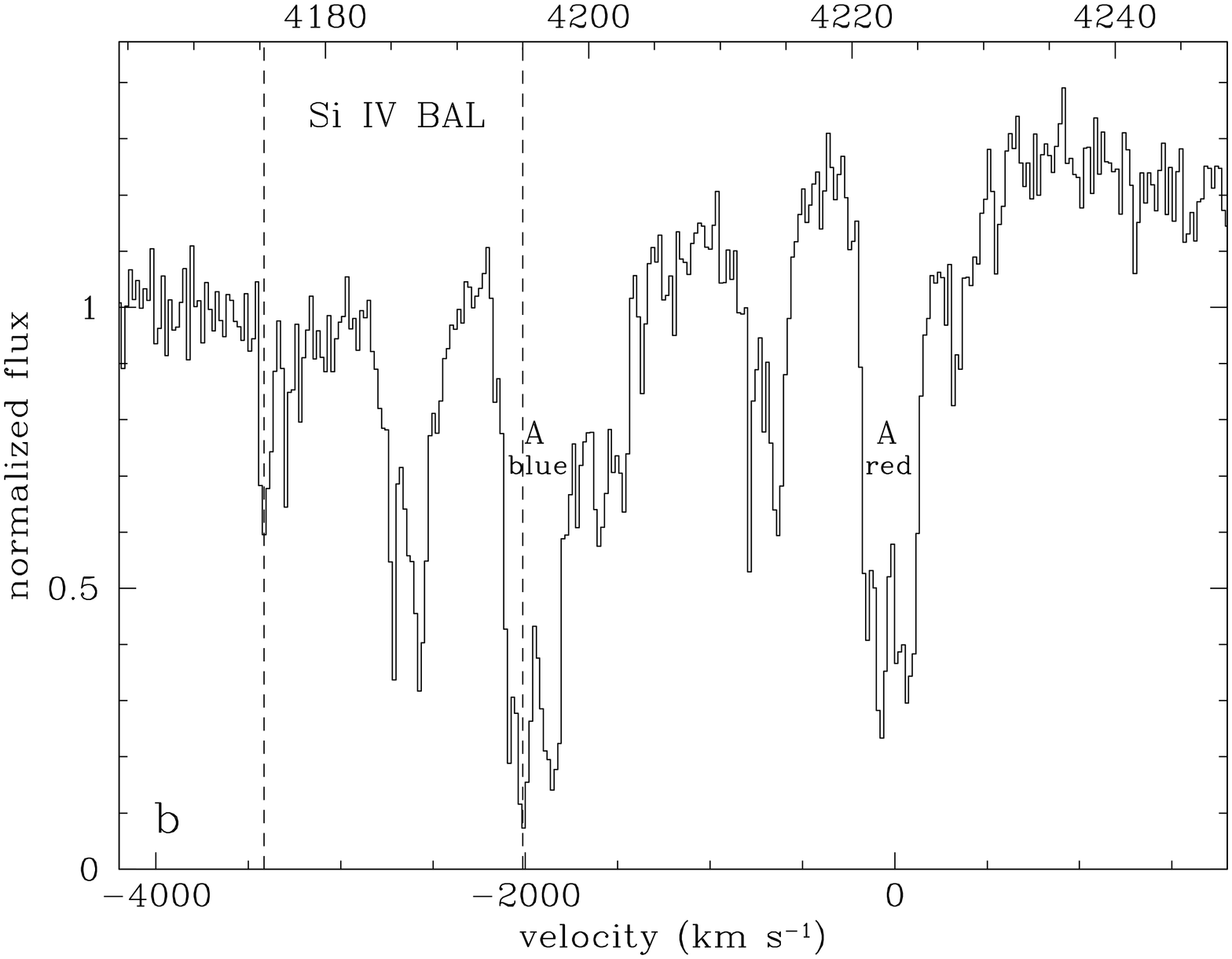,angle=-90,height=6.0cm,width=9.cm}
  \end{minipage}\hfill
  \begin{minipage}[t]{9.cm}
\hspace*{9.cm}
    \psfig{file=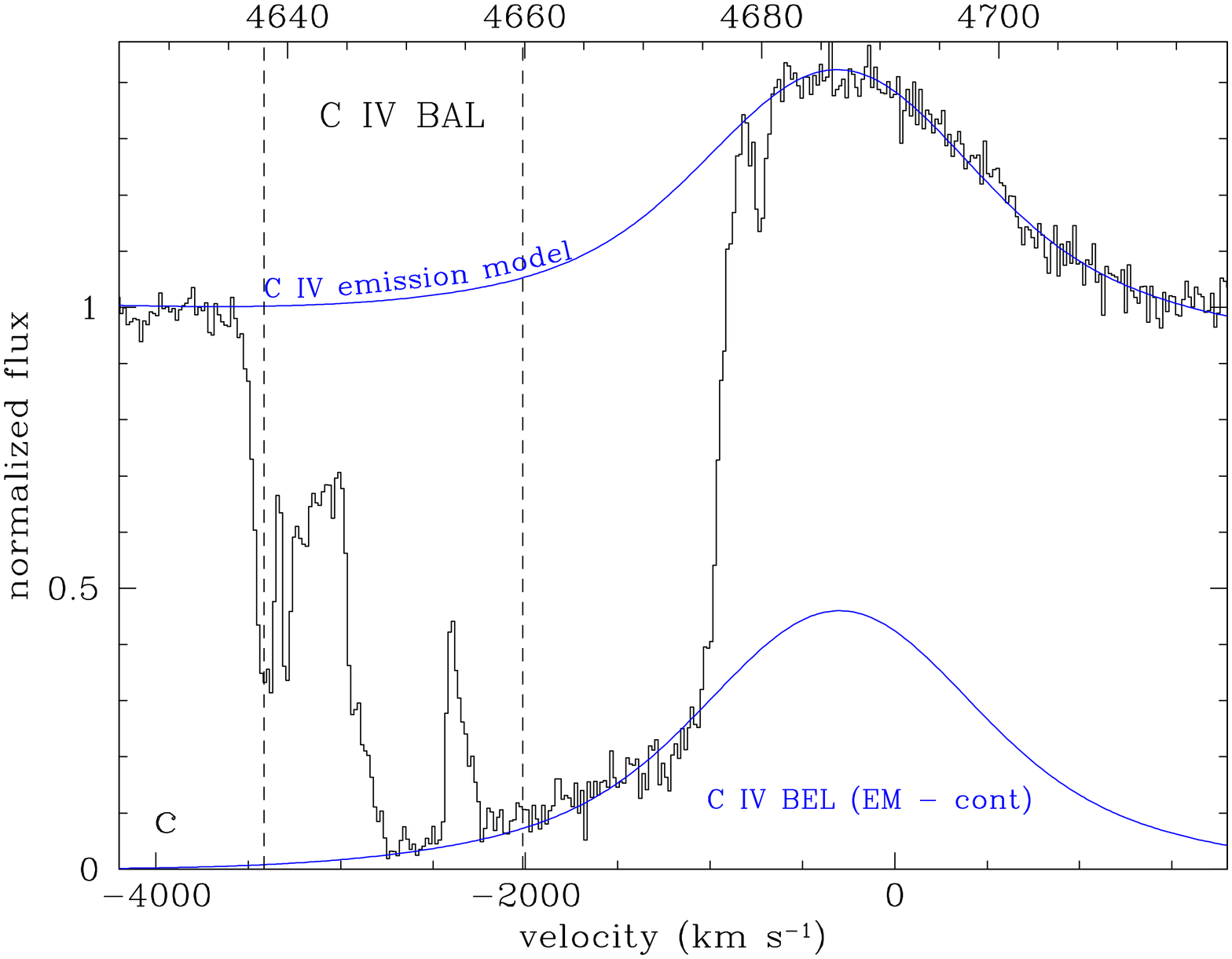,angle=-90,height=6.0cm,width=9.cm}
  \end{minipage}\hfill
  \begin{minipage}[h]{8.5cm}	
%  \begin{minipage}[b]{17.cm}	
  \vspace*{-20.cm}
\hspace*{10.cm}

\caption{Data for the three observed BALs are shown on the same
velocity scale where zero velocity is fixed at the (redshifted)
wavelength of the stronger transition for each line. The top x-axes
show the observed wavelength for each data segment.  The two dashed
lines demonstrate the velocity coincidence of two distinctive
absorption features within the troughs.  In panel $a$, $\Delta d$
gives the expected separation between the two prominent transitions of
\cii\ 1335 \AA\ and shows that the two observed troughs are indeed due
to \cii. In the \siiv~$\lambda\lambda$ 1394, 1403 BAL we observe three
troughs which are seen in both the blue and red components of the
doublet.  We labeled the deepest one as trough A and marked the
position of both its blue and red components.  In panel $c$ we plot
the unabsorbed emission model for \civ~$\lambda\lambda$ 1548, 1551 and
the model for the broad emission line (BEL), which is obtained by
subtracting the continuum from the full emission model.  The excellent
match between the BEL profile and the shape of the deepest trough,
strongly suggests that in this object the BAL flow does not cover the
BEL region.}
\end{minipage}
\end{figure}

\clearpage

\subsection{\siiv\ BAL}

Three distinct troughs are seen in the \siiv\ BAL (Fig.~1b). Since the
total absorption width is only slightly greater than the \siiv\ doublet
separation ($\sim2000$ \km), the two main troughs are seen in both the
blue and red components of the doublet and are unblended with other
absorption.  The ability to measure unblended features from two lines
of the same ion allows us to solve separately for the effective
covering factor and the real optical depth
\cite[]{barlow97,hamann97a,arav99a}.  The effective covering-factor
($C$)\footnote{We do not use the notation $C_f$ (introduced by Barlow
and Hamann) in order to reserve the use of a subscript to
differentiate between continuum and BEL covering factors.}  is defined
such that $(1-C)$ accounts for photons that arise either from regions
not covered by the BAL flow or those that are scattered into the
observer's line of sight.  If scattering into the line of sight is
negligible, then $C$ is the total emission-covering-fraction of the
BAL flow.  In \siiv~$\lambda\lambda$ 1394, 1403 the expected intrinsic
optical depth ratio is 2:1 since the oscillator strength of the
$\lambda$1394 line is twice that of the $\lambda$1403 line.  The
relationships between the residual intensity in the red and blue
doublet components ($I_r$ and $I_b$, respectively), $C$ and the
optical depth are given by:
\begin{eqnarray}
I_r&=&(1-C)+Ce^{-\frac{1}{2}\tau} \\
I_b&=&(1-C)+Ce^{-\tau}, 
\end{eqnarray}
where $\tau$ is the real optical depth of the stronger transition.

We concentrate our analysis on the deepest of the \siiv\ troughs ($A$
in Fig. 1b). As we demonstrate below, we obtain a lower limit to the
true optical depth and minimize the role of the covering factor if we
assume that the flow covers the broad emission line (BEL) region to
the same extent that it covers the continuum source.  
  The first step is to fit an emission
model for the whole \siiv\ region.  We then divide the data by the
emission model to obtain the normalized residual intensities.  Working
in $\log(\lambda)$ space (in which the doublet separation is constant)
 we shift the absorption due to the red 
component by the doublet separation to obtain a dataset which
contains $I_r$ and $I_b$ on the same $\log(\lambda)$ scale.  For each
$\log(\lambda)$ bin we solve equations (1) and (2) for both $C$ and
$\tau$.  The results are shown in Figure 2, where for clarity we
transformed the x-axis to a velocity presentation.  Physical solutions
for equations (1) and (2) exist only if $I_r \geq I_b \geq I_r^2$
\cite[]{hamann97a}. Values outside this constraint are due to photon
shot noise or systematic errors.  Whenever we encountered a bin in
which $I_r \leq I_b$, we treated it as though $I_r= I_b$, i.e.,
$C=1-I_b$ and $\tau=\infty$.  For the segment we have solved for, this
situation arises only once ($\sim -1940$ \km).

\begin{figure}[t]
  \vspace*{-0.6cm}
  \begin{minipage}[t]{9.cm}
\hspace*{4.cm}
    \psfig{file=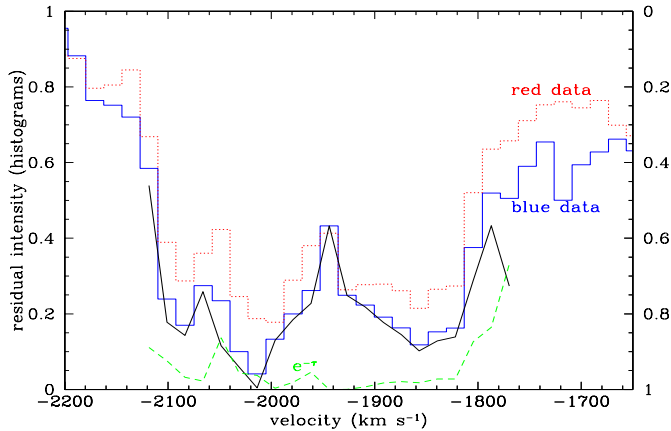,angle=-90,height=6.0cm,width=9.cm}
  \end{minipage}\hfill
  \begin{minipage}[b]{15.cm}	

\caption{ Covering factor and optical depth solutions for the deepest
\siiv\ trough ($A$). The histograms give the residual intensity of the
blue and red components of the trough, where the red data were shifted
by the doublet separation for a direct comparison with the blue data.
The thick solid line shows the effective covering factor ($C$, right y
axis) and the dashed line is $e^{-\tau}$ (plotted on the left y axis),
where $\tau$ is the real optical depth for the blue doublet component.
It is clear that the shape of the trough is largely determined by the
value of the covering factor rather than by variations in the
intrinsic optical depth. }

\end{minipage}
\end{figure}

Figure 2 shows that the covering factor has almost the exact same
shape as $I_b$.  The dashed line shows $e^{-\tau}$, which would have
been the shape of the absorption trough if the coverage were complete.
Since $C(v)$ is almost identical to $I_b(v)$ while $e^{-\tau(v)}$ does
not correlate with $I_b(v)$ we conclude that the shape of absorption
trough $A$ is determined by variation in the covering factor and not
by changes in the real optical depth.  This characteristic is most
noticeable  in the ``hump'' 
between $-2000$ and $-1850$ \km.  If the shape of this hump was
determined by changes in real optical depth, we would expect
$e^{-\tau(v)}$ to mimic $I_b(v)$.  From Figure 2 this is clearly not
the case. In fact the
highest residual intensity is actually the point of largest optical
depth.  The real optical depth across trough $A$ is 3--6 times larger
than the {\it apparent} optical depth ($\tau_{apparent}\equiv
-ln(I_b)$), demonstrating the unreliability of extracting column
densities from measurements of $\tau_{apparent}$.
 
A similar result is obtained for the
second deepest \siiv\ trough (the last trough being too shallow and
partially blended cannot be used for this analysis).  In \S~3.3 and \S~4 we
combine the dominance of the covering factor in determining the shape
of trough $A$ with the information gathered from the \cii\ BAL
(\S~3.3) to produce a geometrical picture for the flow.

As we discuss in the next section, the \civ\ data suggest that the BAL
flow does not cover the BEL region in this object.  If this is the
case for the \siiv\ absorption, how does it affect the results shown
in Figure 2?  The contribution of the \siiv\ BEL to the total emission
is larger for the red doublet component of trough $A$ than for the
blue component, and when we subtract a modeled \siiv\ BEL from the
data the residual intensity of the two troughs are identical within
the noise. In such a case, the lines must be highly saturated (with no
useful upper limit for $\tau$ possible) and the shape of the trough is
determined solely by the behavior of the covering factor.  In \S~4 we
argue that this picture is the simplest interpretation of the data.

\subsection{\civ\ BAL}

For the \civ\ BAL (Fig. 1c) we cannot use the same solution technique
since the intrinsic doublet separation is only 500 \km, much smaller
than the width of the flow ($\sim 2000$ \km), thus the trough is a
blend of the two doublet components.  However, we have an independent
indicator for non-black saturation in this BAL as well.  We model the
unabsorbed emission with a BEL on top of a linear continuum (Fig.~1c).
For the BEL, we used a two Gaussian model that gave an excellent fit
to the unabsorbed part of the \civ\ BEL.  We show the \civ\ BEL (which
is derived by subtracting the continuum from the full emission model)
on the same plot. The flux as a function of velocity at the deepest
part of the \civ\ BAL ($-2700$ to $-1500$ \km) is remarkably similar
to the flux of the modeled \civ\ BEL.  From this we deduce that the
BAL flow in this object does not cover a significant fraction of the
BEL region.  A similar behavior is seen in Q1413+113
\cite[]{turnshek88b}.  Accepting this assertion leads to the
conclusion that the \civ\ BAL flow is optically thick, since it blocks
virtually all the continuum emission.  Therefore, the shape of the
\civ\ BAL trough contains information about the geometry and
kinematics of the flow but not about the column density of the
absorber.

Low resolution data of the \Ly\ BEL (for which we do not have Keck
HIRES coverage), were taken with the KAST double spectrograph at Lick
observatory.  These data (shown in \nocite{white99} White et al.\
1999) support our assertion that the BAL flow does not cover the BELs.
Our data show unambiguous \Ly\ BAL absorption on the blue wing of the
\Ly\ BEL. Between $-1000$ \km\ and $-2200$ \km\ the shape of the
trough is consistent with a covered continuum and an uncovered
BEL. Since the peak flux of the \Ly\ BEL is roughly  twice as
strong as the continuum, and six times stronger than the \civ\ BEL, we
expect to see substantial \Ly\ emission peeking through the BAL flow.
This is indeed the case, for example, at $-1500$ \km\ where the
observed flux is 1.5 times higher than the continuum level, in
agreement with our predictions. In addition to the \Ly\ BAL we also see the
\nv\ BAL in the low resolution data. The shape of this trough also supports
our assertion that the BAL flow does not cover the BELs.

\subsection{\cii\ BAL}
 Absorption associated with the BAL flow is clearly detected in
\cii~$\lambda$1335 (Fig.~1a).  This line is a triplet with components
at 1335.708 \AA, 1335.663 \AA\ (both from an excited level) and
1334.532 \AA\ \cite[]{verner96}.  The 1335.663 \AA\ component is only
11\% as strong as the 1335.708 \AA\ component and is separated from it
by only $\sim$ 10 \km.  For our purposes we can therefore treat the
whole line as a doublet with components at 1335.703 \AA\ and 1334.532
\AA, which have an intrinsic optical depth ratio of 2:1, respectively.
From the comparable absorption equivalent widths seen in
\cii~$\lambda$1334.532 and \cii$^*~\lambda$1335.703, and taking into
account the possibility of saturation, a lower limit of $\sim10$
cm$^{-3}$ can be obtained for the number density of the gas
\cite[]{wood97}.  This lower limit cannot be taken as evidence for the
intrinsic nature of the absorption, since similar number density
values are inferred for some intervening absorption systems \cite[for
example Q1037--2704;][]{lespine97}.  Due to the lower S/N and the
shallowness of the absorption in \cii, we cannot get meaningful
results from trying to solve for the covering factor and real optical
depth in this line.  Even so, the data are strongly suggestive of
saturation since the troughs have an apparent optical depth ratio
less than 2:1.

From Figure 1, it is evident that the \cii\ absorption is perfectly
aligned with the deepest subtrough of \siiv\ trough $A$.  No
significant \cii\ absorption is seen associated with the low-velocity
subtrough of this feature even though its residual intensity in \siiv\
is almost identical to that of the deepest subtrough. (There is also a
third subtrough around $-2090$ \km, but since it is narrower and less
distinct we ignore it.) Our explanation for this occurrence is that
what we see are two distinct outflows.  One outflow might have a lower
ionization equilibrium and thus shows \cii\ absorption.
Alternatively, the flow that shows \cii\ absorption might be in a
similar ionization equilibrium but have a significantly larger optical
depth in all lines, which allows a detection of a small \cii\
contribution.  This picture agrees well with our inferences from
the \siiv\ analysis.  We know that the shape of trough $A$ is
determined by changes in the covering factor, which shows two distinct
subtroughs.  The simplest way to explain one such subtrough is to
assume that an accelerating outflow moves in and out of our line of
sight \cite[]{arav_ghost,arav99a}.  Two such outflows which happen to
cross our line of sight at similar radial velocities will give rise to
the two subtroughs seen in trough $A$.  Since these are not physically
connected it is less of a surprise to detect \cii\ absorption in only
one of them.

\section{DISCUSSION}

To relate our findings to the whole class of BALQSOs, we need to
establish the relationship of the absorption seen in BALQSO 1603+3002
to the BAL phenomenon in general.  \cite{Weymann91} defined a BAL as a
continuous absorption of at least 10\% in depth spanning more than
2000 \km, discounting absorption closer than 3000 \km\ bluewards of
the emission peak.  In Figure 1c we show the data for the \civ\
BAL. The width of continuous absorption deeper than 10\% is 2600 \km,
which satisfies the width criterion, but most of the absorption is at
velocities closer than --3000 \km\ from the emission line peak.
However, the --3000 \km\ condition was introduced in order to
unambiguously distinguish between associated absorbers and ``classical
BALs'' but does not hold any physical meaning.  The flow in BALQSO
1603+3002 shows non-black saturation and the \civ\ data suggest that
the flow does not cover the broad emission line region of the object
\cite[with a size of $\sim0.1$ pc.;][]{netzer90}.  Each of these 
independent findings mark the flow as arising from the vicinity of the
central source and as being physically similar to ``classical BALs''.
With the data improvements available in recent years (especially high
resolution spectroscopy) we advocate classification of absorption
systems based on their physical characteristics
\cite[see][]{barlow97}, rather than the older phenomelogical one.

The geometry that we proposed for trough $A$ (\S~3.3) can be extrapolated
to the full observed BAL.  We have already mentioned that the
structure seen in the trough situated at $\sim-2600$ \km\ (see Fig 1b)
is due to variations in the covering factor.  Therefore, following the
arguments we used for trough $A$, it seems plausible that this trough
is also the result of two outflows that cross our line of sight at
similar radial velocities.  If we extend this picture to the
shallowest trough at $\sim-3400$ \km, which also shows two adjacent
absorption features, we are led to postulate six different outflow
components in the full BAL.  Multi-component flows might be quite
common in BALQSOs since many of them show several absorption troughs.
For example, in the spectra shown by \cite{korista93} there are
four \civ\ troughs in Q0146+0142, 3 in Q0226--1014, 3 in Q0932+5010
and 4 in Q2240--3702.  One unexplained coincidence in our flow model
is the occurrence of three pairs of closely adjacent subflows.
Starting from trough $A$, the separations between the deepest
absorption features in each trough are: 154 $\pm 8$, 136$\pm 8$ and
117$\pm 8$ \km.  Having three such absorption pairs all with
separations between 100--150 \km, across a full velocity width of more
than 2000 \km\ seems improbable without a dynamical justification.

 Independent of our flow model, however, solving for the \siiv\ doublet
components shows that the structure in at least the first two troughs
is mainly due to changes in the covering factor (see Fig.~2).
Assuming that the flow covers the whole emission region leads to
$\tau_{real}$ values between 2--5 across trough $A$.  Alternatively,
assuming the flows do not cover any appreciable part of the BEL region
yields indistinguishable residual intensities for both the red and blue
components of each trough (after subtracting the BEL contribution).
  In this case $\tau_{real}$ values are
between $5-\infty$.  The latter option seems more physical for two
reasons.  First, from the \civ\ data we infer that the flow as seen in
\civ\ does not cover the broad emission line.  There is no reason to
assume that the \siiv\ case is different.  Second, in the absence of a
physical preference for $\tau_{real}$ values of order unity, values
between 2--5 necessitates some fine tuning whereas the range
$5-\infty$ is simply much more probable numerically.  

Although the three flow components are seen in both \civ\ and \siiv,
there are important differences between these two manifestations.  The
\civ\ absorption is always deeper and somewhat wider than the \siiv\
flow.  Also, in \civ\ there is no trace of the large variations
in covering factor seen in \siiv, trough $A$.  These differences show that in
the two main troughs the covering factor is ion dependent.  A model
based on column density gradient and kinematic effects can explain
this behavior qualitatively \cite[]{arav99a}.

\section{SUMMARY AND CONCLUSIONS}

High resolution spectroscopy of BALQSO 1603+3002 has yielded important
diagnostics for the nature of quasar outflows.  The presence of two
relatively wide but still unblended doublet components of \siiv\ in
its spectrum has allowed us to distinguish between the effects of
column density and covering factor in determining the shape of the
absorption troughs.  A straightforward solution of equations (1) and
(2) demonstrates that changes in the covering factor are responsible
for the troughs' shape as opposed to variations in the real optical
depth. This result was independently supported by the findings from
the \civ\ BAL, which indicated that the flow does not cover an
appreciable portion of the BEL region (Further evidence for the non-covering
comes from the low resolution data of the \Lya\ and \nv\ BALs;
see \S~2.3.).  Subtracting the BEL contribution, the resultant BAL is
black across a considerable span and therefore saturated.

The inference from the \civ\ BAL, that the BEL region is not covered
by the flow, strengthens the results derived from the \siiv\ analysis.
After subtracting a modeled \siiv\ BEL, the residual intensities of
the blue and red components of the troughs are identical within the
noise.  Such occurence indicates that the absorption is highly
saturated and that the shape of the trough is solely determined by
changes in the covering factor.  It also suggests that the transition
from opaque matter to $\tau \ll 1$ is quite sharp.  Once we know that
the shape of the absorption line is due to the covering factor, it is
natural to model the structure within trough $A$ of \siiv\ as arising
from two separate outflows.  Independent evidence for this assertion
comes from the \cii\ BAL which shows an absorption feature which
coincides with only one of the subtroughs seen in \siiv, trough
$A$. This result supports a picture of a BAL region consisting of
several flows that appear to have different properties, either as a
result of a different ionization state or simply because of different
optical depth.  However, the real situation must be more complicated
since the \civ\ and \siiv\ BALs show different covering factors at the
same velocities.

Our findings have important implications for abundance studies of the
flows.  As we showed, extracting $\tau$ from the depth of the trough
using $\tau=-ln(I_r)$ severely underestimates the true optical depth.
This leads to a similar underestimation of the resultant column
density.  Since abundances are determined by a {\it relative}
comparison of column densities after accounting for the ionization
equilibrium, underestimating the hydrogen column density can produce
erroneously high absolute abundances for all the heavy elements.
Differential metal abundance determinations are also susceptible to
large errors arising from underestimating column densities.  Based on
the apparent column densities in Q0226-1024, \cite{turnshek96} found
(their table 4, first model) that Si and S are highly enriched
relative C, compared to solar ratios: (Si/C)$\simeq4$(Si/C)$_{\odot}$,
(S/C)$\simeq8$(S/C)$_{\odot}$.  Similarly, \cite{junkkarinen97} found
(P/C)$\simeq60$(P/C)$_{\odot}$ in PG~0946+301.  When compared to the
solar abundance ratios, (C/Si)$_{\odot}$=11, (C/S)$_{\odot}$=20 and
(C/P)$_{\odot}$=1000 \cite[]{grevesse89}, a trend of higher
enrichment for rarer elements is evident.  This surprising and suspect
correlation can be eliminated if we accept that the column densities
are large, the absorption is saturated, and consequently the shapes of
the troughs are only mildly dependent on the real optical depth.  In
such a case, we would expect only a small variation in the depth of
troughs which arise from different elements, even when the abundances
differ by large factors.  If one does not assume saturation, a
progressively higher enrichment for rarer elements has to be invoked
to explain the small variation in apparent column density.  Non-black
saturation accounts for this without invoking fantastic metal
enrichment.  Rare elements are simply less saturated than more
abundant elements\footnote{\cite{turnshek96} also found higher
enrichment relative to carbon for nitrogen and oxygen.  These findings
can also be explained by the saturation scenario.  If we assume that
all the BALs in Q0226-1024 are similar in depth and shape (which is
correct to within a factor of 2), we deduce higher apparent column
densities for lines with weaker oscillator strength.  This is exactly
the case for the \oiii, \oiv, and \niii\ BALs observed in Q0226-1024,
and the high column densities deduced for these ions \cite[Table
2]{turnshek96} are largely responsible for the very high enrichment
reported for these elements.}.

Based on the results shown
in this paper and on independent evidence for non-black BAL saturation
(see \S~1), we conclude that BAL abundances claims in the literature
which are based on {\it apparent} $\tau$ should be treated
with the utmost caution.

\section*{ACKNOWLEDGMENTS}
We thank the referee Kirk Korista for several valuable suggestions.
Part of this work was performed under the auspices of the US
Department of Energy by Lawrence Livermore National Laboratory under
Contract W-7405-Eng-48. We acknowledge support from the NRAO, NSF
grant AST-9802791, STScI and Sun Microsystems.

\pagebreak

%\bibliography{references}

%\bibliographystyle{natbib}

\end{document}